\documentclass[twocolumn,aps]{revtex4-2}
\usepackage{epsfig}
\usepackage{amsmath}
\usepackage{soul}

\begin{document}
\title{100-ns-level timing holdover after 12 years for rubidium atomic fountains}

\author{Steven~Peil$^1$}
\email{steven.e.peil.civ@us.navy.mil}

\author{T.~G.~Akin$^1$}

\author{J.~D.~Whalen$^{1,2}$}

%\author[1]{Steven~Peil}
%\author[1]{T.~G.~Akin}
%\author[1,2]{J.~D.~Whalen}
%\affil[1]{Precise Time Department, United States Naval Observatory, Washington, DC 20392}
%\affil[2]{Computational Physics, Inc., Springfield, VA}

\affiliation{$^1$Precise Time Department, United States Naval Observatory, Washington, DC 20392}
\affiliation{$^2$Computational Physics Inc., Springfield VA 22151}

\date{\today}

\begin{abstract}

While atomic frequency standards are improving at a staggering pace, the timing community has relied on the same continuously running atomic clocks for decades: commercial cesium beams and hydrogen masers. Challenges in incorporating the latest technological advancements into operational clocks has resulted in technology lag compared with frequency standards that consequently impacts timing applications, such as system synchronization, positioning and timescales. The first cold-atom clocks to contribute to the free running international atomic timescale, EAL, are the four rubidium fountains in operation at the U.S.~Naval Observatory in Washington, DC, that came online in 2011.  With 12 years of uninterrupted data from the International Bureau of Weights and Measures (BIPM) from Modified Julian Date (MJD) 56074 to MJD 60429, we report on the long-term timing performance of these clocks. The highest performing fountain exhibits TDEV of 8~ns at $\sim 3$~years and a holdover of BIPM's best timescale of $\pm14$~ns at 12 years.  

\end{abstract}
%\pacs{03.65.Ta, 03.65.-w, 03.65.Yz}

\maketitle

\section{Introduction - Frequency (Standards) and Time (Keeping)}

\begin{figure}
\includegraphics[width=0.45\textwidth]{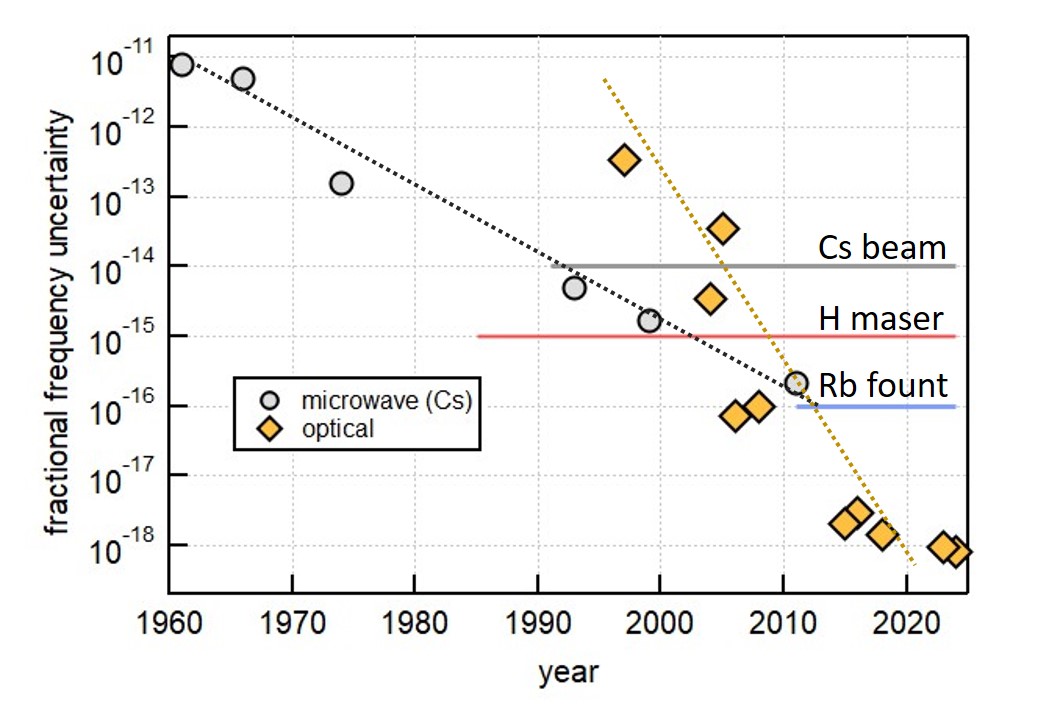}
\caption{Plot of demonstrated frequency uncertainty for microwave (grey circles) and optical (gold diamonds) frequency standards. The three horizontal lines represent the approximate instability floor of operational clocks, which are represented as stagnant over the years. Frequency uncertainties are obtained from references~\cite{PhysRevLett.133.023401, PhysRevLett.123.033201, mcgrew, PhysRevLett.116.063001, nicholson, ludlow_sr, PhysRevLett.97.020801, katori, npl_sr_ion, hansch} (optical) and~\cite{Heavner_2014, guena, Jefferts_2002, J_H_Shirley_2001, 4314340, 4313501} (microwave). The operational clock floors are from manufacturer datasheets for masers and cesiums~\cite{datasheets}, and from Ref.~\cite{Peil_2016} for the rubidium fountains.
}\label{f.moore}
\end{figure}

Frequency is the most precisely measured physical quantity, with state-of-the-art frequency standards capable of measurements with fractional systematic uncertainties below $10^{-18}$~\cite{PhysRevLett.123.033201, PhysRevLett.133.023401} and instabilities below $10^{-20}$~\cite{Bothwell2022}.  This astounding precision is 5000 times better than that achievable 30 years ago, when the accuracy of (laboratory) microwave cesium-beam standards plateaued at $5\times 10^{-15}$~\cite{J_H_Shirley_2001}.  Since then, a Moore's-law type of improvement in frequency standards has occurred (see Fig.~\ref{f.moore}), enabled first by the adoption of cold atoms~\cite{RevModPhys.70.721} and subsequently by the move to optical frequencies made possible by the development of frequency combs and narrow line-width lasers~\cite{RevModPhys.87.637}. Presently, record stabilities and uncertainties are achieved using optical-lattice frequency standards, in which $\sim 10^4$ neutral atoms contribute to the clock signal~\cite{RevModPhys.83.331}. Advances in frequency metrology are ushering in the field of relativistic geodesy~\cite{Flury_2016} and will lead to the redefinition of the second in the International System (SI) of Units~\cite{fritz}.  

Time, at the most precise operational level, is measured by clocks that track the phase (integral of the frequency) of a periodic oscillator, such as a frequency standard.  Interruptions in the frequency standard irretrievably perturb the timing signal, which necessitates another, more reliable, holdover clock to maintain timing.  While an optical lattice can serve as the basis of a clock that does not lose a second over $10^{10}$ years, these highly complex systems typically do not operate continuously, without user intervention, for more than several days or weeks~\cite{Kobayashi_2020}. For applications requiring uninterrupted timing, such as navigation, sensor fusion and network synchronization~\cite{ion}, reliable clocks that operate continuously and with long life are utilized to provide continuous timing and holdover between calibration events with complex frequency standards or other external references.

While ultimate accuracy is the most important property of a frequency standard, accumulated timing error is the defining metric for a clock. %Timing error accumulates when a clock synchronized to a superior reference runs freely for an interval,
When a free running clock is trying to maintain the time of a superior reference to which it had been synchronized, it can be said to be operating in ``holdover''~\cite{4638900,NPL}. The timing error of a clock in holdover is often specified for intervals from days to months, but stringent holdover capabilities for long intervals are important for applications where independence from external sources of time is required. Coordinated Universal Time (UTC), the international time standard, is broadcast to the world via the Global Positioning System (GPS) and other Global Navigation Satellite System (GNSS) networks and is communicated to the world's timing and metrology labs via the monthly Circular T report from the Bureau of International Weights and Measures (BIPM, Bureau International des Poids et Measures)~\cite{circulart}. Fifteen of the eighteen critical infrastructure sectors designated by the U.S Department of Homeland Security rely on time provided by GPS~\cite{ion}. For purposes of redundancy and resilience, users of precise time benefit from the ability to be independent of GNSS~\cite{9583423,ion} --- and even the BIPM --- for extended periods in case of unforeseen disruptions. In the near future, UTC may need to be transmitted to the lunar surface or lunar orbit~\cite{BAKER2024} (and in the more distant future, to other solar system bodies) to maintain some degree of interoperability between terrestrial and cislunar systems, and the geometric (and other) constraints on time transfer in this theater suggest that long periods of holdover will be necessary. 

The most stringent timing applications involving navigation and surveillance  require timing at the nanosecond level~\cite{ion}. The power grid and communication systems currently require microsecond level timing~\cite{ion, 10093870}, but timing requirements are only expected to become more stringent. GNSS can transfer time with 5-10~ns accuracy, and timing labs can remain within 1~ns of UTC, but there are currently no long-term holdover capabilities that can sustain such precision. Commercial cesium beams and hydrogen masers can provide 
ns-level holdover for $\tau$ of a day and a week respectively, but timing error grows as $\sqrt{\tau}$ in the best case scenario for cesium beams and grows more aggressively for masers. 

Four rubidium fountains at the U.S. Naval Observatory (USNO) in Washington, DC operate as continuous clocks rather than frequency standards, the traditional role of atomic fountains. In operation for over 13 years, these clocks have contributed without interruption to Coordinated Universal Time (UTC) for 12 of those years, providing continuous phase records that enable timing error and holdover assessments over these unprecedented epochs. In this work we use data reported by the BIPM from Modified Julian Date (MJD) 56074 to MJD 60429 to highlight the reliability, stability and holdover capabilities of the USNO rubidium fountains in comparison to other operational clocks.

\section{Clock Technology and Rubidium Fountain Clocks}

As illustrated in Fig.~\ref{f.moore}, the technology for continuously operating clocks has not kept pace with that of frequency standards; in fact, commercial clock technology has been somewhat stagnant for several decades. Most operational clocks in use today are commercial cesium beams~\cite{Cutler_2005} and hydrogen masers~\cite{Vessot_2005}, with current models similar to those circa 1990. The lag in precision of clocks compared to frequency standards stems from the challenge of incorporating key technological advances into robust, continuous, user-free devices. Neither laser cooling nor optical spectroscopy made their way into operational clock technology until fairly recently.  Presently, several optical clocks using vapor cells~\cite{optical_sea} and cold-atom microwave clocks are being sold~\cite{drip}, but reliability and long-term frequency stability of these devices will take years to fully assess.  

On the other hand, the technology for these commercial clocks is extremely mature, and the clocks reliably output a timing signal continuously for years. High-performance commercial cesium beams typically run for 5 years before needing a cesium refresh, and hydrogen masers can run maintenance-free for even longer, usually limited by ion-pump saturation. It is this continuous, reliable operation over many years that distinguishes an operational clock from a frequency standard, which is designed to reach ultimate accuracy with less emphasis on reliability and availability. 

One of the first introductions of cold atoms into operational clocks was the ensemble of rubidium fountains used for timing at USNO~\cite{Peil_2016}, which, unlike all other atomic fountains at the time, operate continuously for long intervals rather than intermittently as frequency standards. Compared to beam clocks, laser cooling enables longer interrogation time and cold, slow atoms have better understood and controlled systematic shifts. This progression from a (thermal) beam clock to a (cooled) fountain clock improved the stability floor for operational clocks to close to $10^{-16}$ (Fig.~\ref{f.moore}), from typical values of $10^{-14}$ for cesium beam clocks and $10^{-15}$ for hydrogen masers.

Despite the increased complexity of using laser cooling, the fountains are exhibiting long lifetime with only modest maintenance required. Atoms are trapped from a room-temperature vapor, extending the source lifetime compared to cesium beam clocks. Additionally, there is no ion pump saturation as occurs in hydrogen masers. As with any similar technology, the laser system in a fountain is the biggest concern for continuous operation and long lifetime. Using mature telecom-based technology for the past 5 years, we observe robust and long-lived laser performance. A well regulated environment with temperature stability of 100~mK and humidity constant to 2\% is vital to maintaining reliable, continuous operation.

Rubidium fountain design and operation have been discussed previously~\cite{2005IFCS,Peil_2016}, and we only provide a brief overview here. The fountains were designed to run as continuous clocks to contribute to the USNO timescale, similar to a cesium beam or hydrogen maser. Because the USNO fountains were not intended to serve as frequency standards, the choice of atom was not constrained to cesium. Rubidium~\footnote{There are several institutions in the world that use rubidium fountains. Rubidium frequency standards are operated at SYRTE~\cite{Guena_2014} and NPL~\cite{Ovchinnikov_2011}, and a continuously running fountain is used at NSTC~\cite{9933887}.} was chosen for technical reasons: the much smaller cold-collision frequency shift~\cite{syrte_cc,  yale_cc}, which reduces the sensitivity to long-term fluctuations in atomic density, and the ability to use telecom laser technology to generate the required 780~nm laser light via second-harmonic generation (SHG). After carrying out preliminary investigations of SHG with nonlinear waveguides~\cite{1275080}, we settled on commercial laser systems using a 1560~nm fiber laser that is amplified and frequency doubled in a single-pass periodically-poled lithium niobate (PPLN) crystal to produce 1~W of fiber-coupled light at 780~nm. The laser light is delivered to the vacuum chamber to create a (1,1,1) magneto-optical trap (MOT) geometry with a magnetic-field gradient of 3~G/cm that collects $\sim10^8$ $^{87}\mathrm{Rb}$ atoms in 250~ms from a room-temperature vapor. Along with the rest of the vacuum system, the trapping region is contained within a set of magnetic shields that create a low-field (1~$\mu$G) environment for molasses cooling.

The output clock signal from a USNO Rb fountain is a steered 5~MHz local oscillator (LO) from which the resonant 6.834~GHz Ramsey drive is generated. This output is measured against USNO's clock ensemble using conventional frequency counter and dual-mixer measurement systems. The fountain performance is limited by contributions to instability from quantum projection noise (QPN) and the LO, a hand-selected quartz crystal oscillator. The LO noise dominates the short-term instability of the clock, yielding a typical Allan deviation, $\sigma_y(\tau)$, of $(1.5 - 2.0)\times10^{-13}/\sqrt{\tau}$. Future upgrades to optical local oscillators or cryogenic sapphire oscillators should enable performance closer to the QPN limit of $5\times 10^{-14}/\sqrt{\tau}$~\cite{Peil_2024, Bothwell:25}.

The LO can be phase locked to a hydrogen maser or other stable reference to ensure the availability of a reliable flywheel. If fountain operation is interrupted the LO frequency is set to the median frequency from the previous hour and no further steers are applied until the nominal operating condition is recovered. As long as the drift in LO frequency over the duration of the interruption is small there is minimal impact to the fountain frequency record. We take advantage of the many low-drift hydrogen masers at USNO and dedicate one to each fountain local oscillator to serve as a flywheel. 

\begin{figure}
\includegraphics[width=0.45\textwidth]{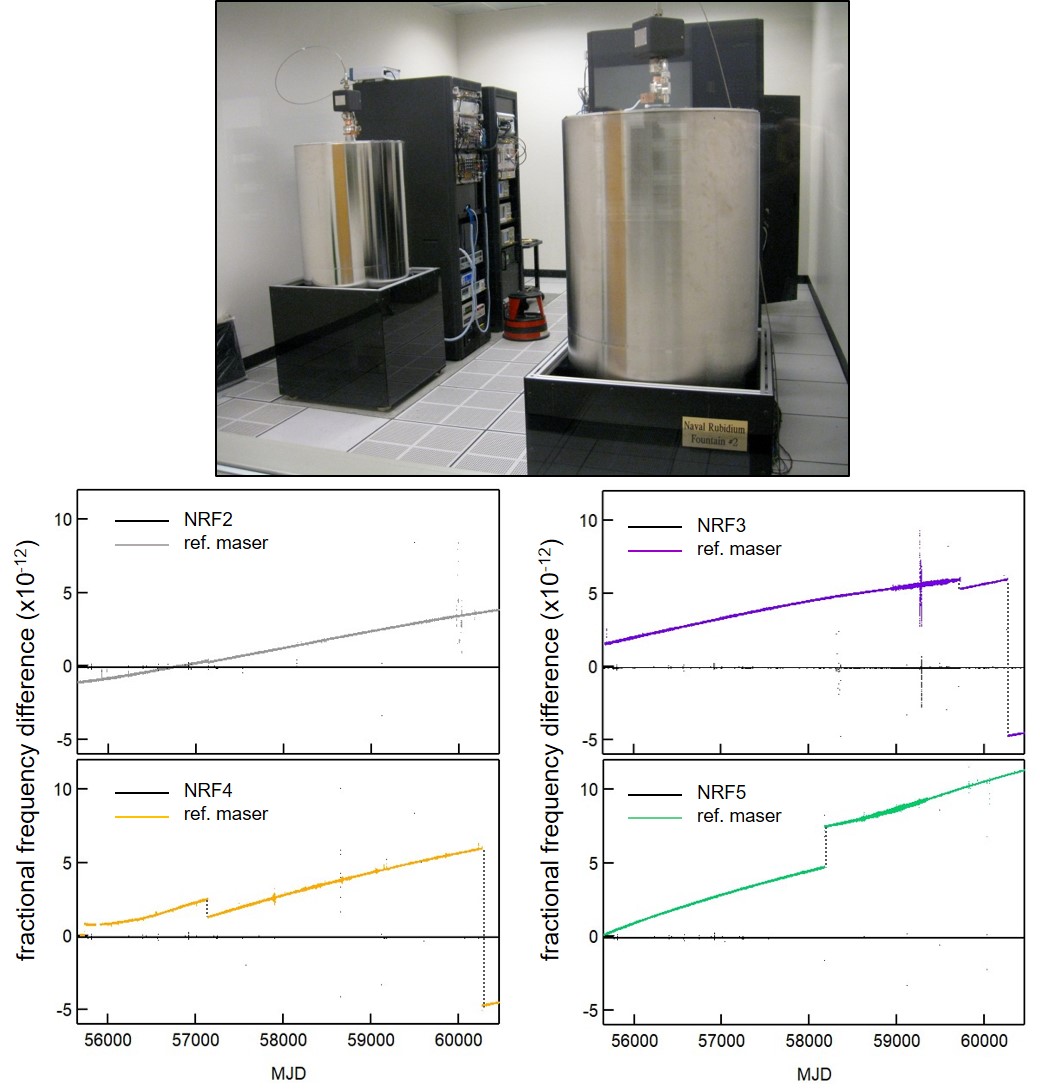}
\caption{Picture of two rubidium fountains (top).
Each system consists of a vacuum chamber enclosed in a set of magnetic shields (front) and two instrument racks for optical and electrical equipment (back).  The hydrogen masers that are available as flywheels are located in a different room.  Plots of the frequency of each fountain against UTC(USNO) as well as each fountain's measurement of its reference maser (bottom). For three of the four fountains, the reference maser needed to be replaced on at least one occasion, as indicated by abrupt changes in the recorded frequency (vertical dashed lines added for clarity). Brief periods of increased noise reflect momentary degradation in either the maser or fountain stability. Fountains are labeled as NRF2, NRF3, NRF4 and NRF5.}
\label{f.fountain}
\end{figure}

All four USNO rubidium fountains in Washington, DC have been in continuous operation since 2011. In Fig.~\ref{f.fountain}, each fountain's frequency measured against UTC(USNO) over these 13 years is shown, along with the frequency of each fountain's reference maser. Abrupt changes in the measured maser frequency correspond to times when the reference maser needed to be replaced due to a saturated ion pump or other performance issue. For three of the four fountains, the reference maser has been replaced at least one time.  Technical issues with the fountains and equipment upgrades to date amount to minor interruptions that are handled gracefully by suspended steering and a stable flywheel; major issues such as depletion of rubidium or saturation of an ion pump would take the clock out of commission. 

\section{Timescales and UTC}
Even with the reliability of the best operational clocks, a weighted ``average'' of clocks, a timescale, is used to ensure that a continuous phase record is maintained for critical applications~\cite{10.1063/1.3681448}. Cesium beams and hydrogen masers have been the primary clocks used in timescales, including UTC, for decades, with the USNO rubidium fountains contributing to UTC for the past 12 years.  

UTC is the authoritative timescale used by the international community and is generated using hundreds of atomic clocks throughout the world~\cite{Panfilo_2019}. It is a paper timescale with physical realizations at most timing and metrology labs, where the physical timescale of lab $X$ is designated UTC($k$).  Timing labs that contribute to the generation of UTC report the value of their clocks' phases, recorded every 5 days with respect to UTC($k$), each month to the BIPM along with high-precision time transfer of UTC($k$). These clock data are used to generate a free atomic timescale, EAL (\'{E}chelle Atomique Libre), by weighting each contributing clock according to its stability and predictability. The clocks contributing to EAL are almost exclusively cesium beams, hydrogen masers, and rubidium fountains (see the Appendix for more on the clock types contributing to EAL). Measurements of certain maser frequencies carried out by primary or secondary frequency standards enable the frequency of EAL to be calibrated to the SI definition of the second, creating International Atomic Time (TAI). UTC is generated from TAI by adding or subtracting leap seconds in order to stay within 0.9~s of Universal Time (UT1), the rotation angle of the Earth with respect to the International Celestial Reference Frame. 

\begin{figure}
\includegraphics[width=0.45\textwidth]{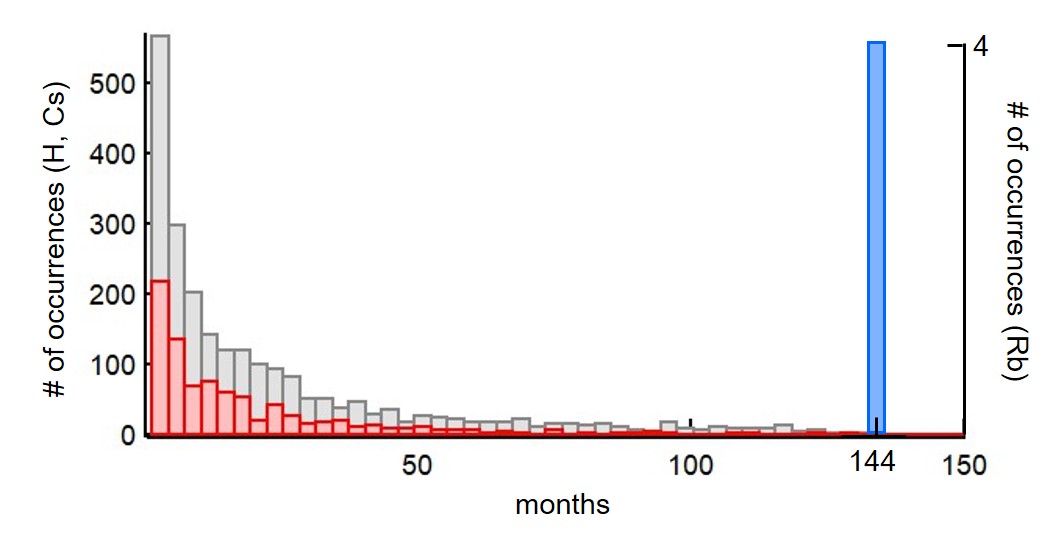}
\caption{Histogram of the length of uninterrupted reporting periods for the three clock types comprising EAL - cesiums (gray, left axis), masers (red, left axis) and fountains (blue, right axis). The $x$-value for the fountain data can be expected to increase.} \label{f.hist}
\end{figure}

Each month the BIPM publishes evaluations of each contributing clock's rate, expressed in nanoseconds lost or gained per day with respect to TAI, its frequency drift, and the assigned weight for generation of TAI~\cite{circulart}. 
If a clock is offline (poor performance or needing maintenance) and is not reported to the BIPM for a given month's assessment and inclusion in TAI, its phase record cannot be continuously tracked. When the clock is reported again, the phase must be tracked anew. The median continuous reporting interval for hydrogen masers is less than 13 months and for cesiums beams less than 18 months; a histogram of the length of continuous reporting intervals for the masers and cesiums considered in this analysis is shown in Fig.~\ref{f.hist} ~\footnote{In this work we analyze only BIPM clock types 35 (High Performance Model 5071A cesium beam) and 40 (unspecified hydrogen maser).  See the Appendix for more on BIPM clock types.}. The USNO rubidium fountains reported and were included in the generation of TAI for every reporting period from MJD 56074 through MJD 60429~\footnote{As of MJD 60777, the fountains have still reported continuously, a total of more than 13 years}, a total of 144 months, and continuous phase records can be tracked for this entire 12 year interval. 

\subsection{Rates and Drifts}

\begin{figure}
\includegraphics[width=0.45\textwidth]{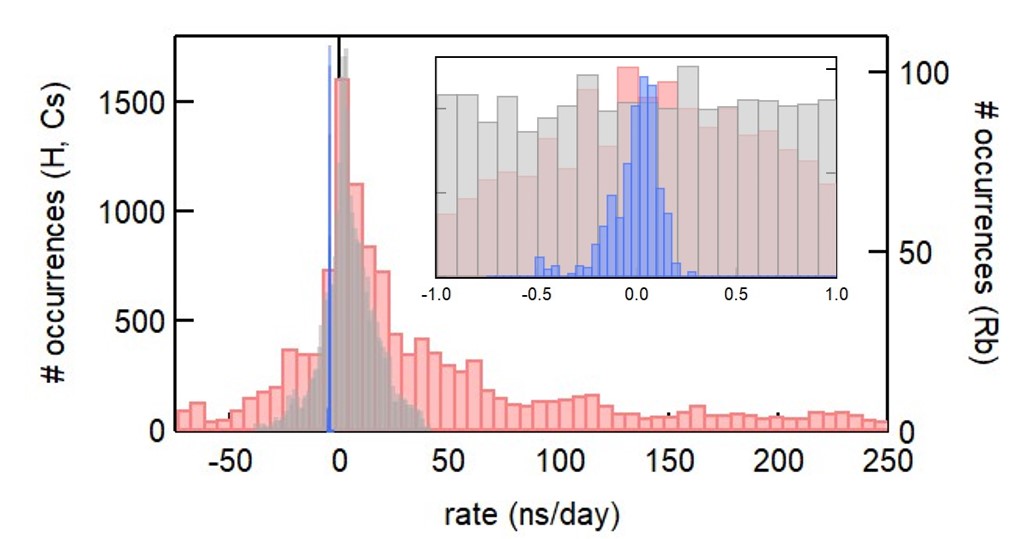}
\caption{Histogram of rates of  cesiums (grey), masers (red), and fountains (blue) with respect to TAI, over 12 years. The units used by the BIPM are (ns/day)$=1.16\times 10^{-14}$. Inset: Re-scaled $x-$axis to show more clearly the distribution of fountain frequencies. The fountain and cesium distributions have been shifted in the inset to enable a better comparison of the spreads in each set of values, and the maser and cesium distributions have been re-binned in the inset.} \label{f.rates}
\end{figure}

Values of a clock's phase over the past month are referenced to EAL to determine the clock's rate, which is subsequently published with respect to TAI. In Fig.~\ref{f.rates}, histograms of the frequencies for cesium beams, hydrogen masers and USNO rubidium fountains measured against TAI are shown. The dispersion of these frequencies for a single type of clock are characterized by standard deviations of $1.2\times 10^{-13}$ for cesiums, $2.3\times 10^{-13}$ for masers, and $1.3\times 10^{-15}$ for fountains.

The fountains are reported to the BIPM without adjusting frequencies for systematic biases with respect to the SI second, resulting in the distribution being offset from zero. The value of this offset is dominated by contributions from the three largest biases: the gravitational redshift ($-8.36\times 10^{-15}$), the blackbody radiation shift ($1.177\times 10^{-14}$), and the second-order Zeeman shift ($-6.495 \times 10^{-14}$). The distribution of fountain frequencies gives a mean frequency offset of $-6.790(5)\times10^{-14}$ with respect to TAI. The inset in Fig.~\ref{f.rates} shows more clearly the distribution of fountain frequencies, after shifting the values so that they are centered on zero. The histograms for masers and cesiums have been re-binned in the inset from the primary graph. 

\begin{figure}
\includegraphics[width=0.45\textwidth]{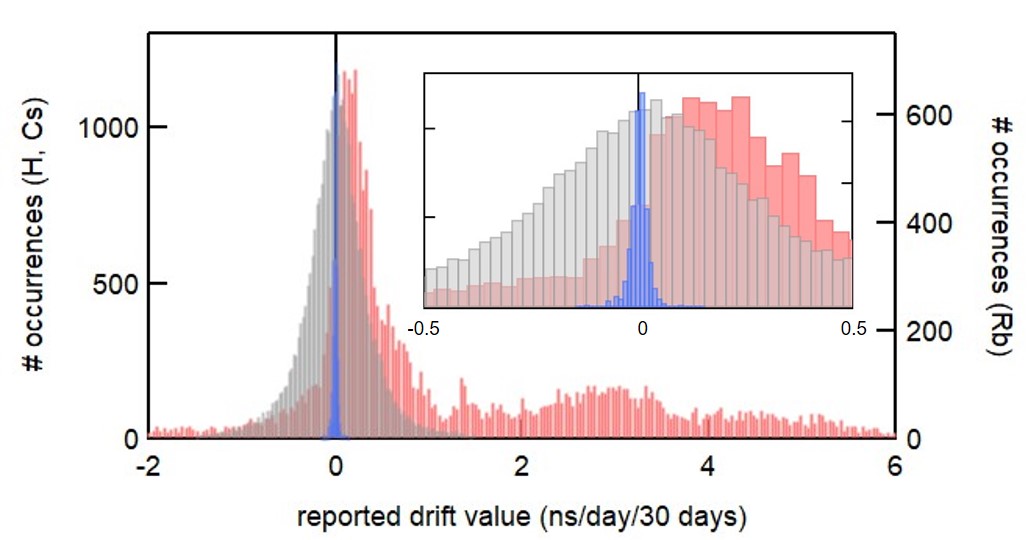}
\caption{Histogram of drift rates for cesiums (grey), masers (red), and fountains (blue), published by the BIPM over 12~years. The rates are averages for the 30-day interval for a given UTC determination, measured using 3 months (currently) or 6 months (before 2017) of clock frequency measurements with a weighted average of the primary frequency standards as a reference. The units used by the BIPM are (ns/day)/30 days $= 3.89\times 10^{-16} /\mathrm{day}$. The region around zero drift is emphasized in the inset.} \label{f.drifts}
\end{figure}

The frequency drift rates for clocks contributing to TAI are measured using the prior 3 months of relative frequency between the clock and a weighted average of the primary frequency standards~\footnote{Before 2017, 6 months of data were used to determine frequency drift.}. A histogram of the drift rates for cesium beams, hydrogen masers, and the USNO rubidium fountains for 12 years is shown in Fig.~\ref{f.drifts}. The distribution for masers shows a primary peak at a nonzero value, with some smaller peaks that seem to be related to maser aging~\cite{PhysRev.126.603, L_Essen_1973}.
The distribution of drift rates for cesiums shows a nonzero mean of $4.0(2.8)\times 10^{-17}/\mathrm{day}$. This is consistent with other values measured informally in the past, but using longer epochs (1.5~years)~\cite{meas_mult_fount, PhysRevA.87.010102_lpi}. The mean for the measured fountain drift rates is $-3.4(4.0)\times 10^{-19} /\mathrm{day}$, consistent with zero. 

\subsection{Weights}

\begin{figure}
\includegraphics[width=0.45\textwidth]{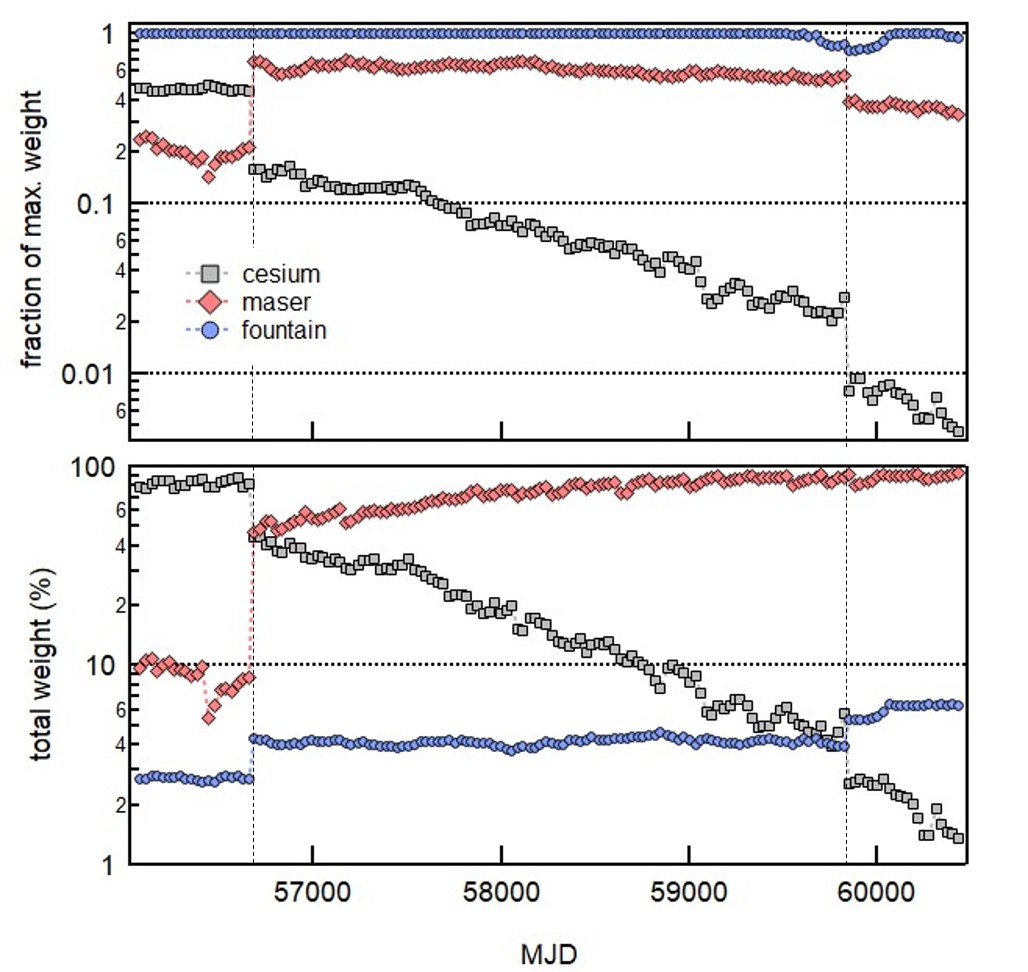}
\caption{The average weight assigned to each type of clock comprising EAL as a fraction of the maximum allowable weight (top). The total weight assigned to each type of clock (bottom). Data shown are cesiums (grey),  masers (red), and fountains (blue). The dashed, vertical lines show when the weighting algorithm or maximum weight was changed as discussed in the text.} \label{f.weights}
\end{figure}

The ALGOS timescale algorithm is used to compute EAL and TAI from the clock data reported to the BIPM~\cite{algos}. Each clock reported is assigned a weight for that month that depends on the behavior and performance of the clock averaged over the previous 12 months. A maximum allowable weight ensures that no clock has outsized influence. The average weights of high-performance cesium beams, hydrogen masers and the USNO rubidium fountains contributing to TAI over 12 years, normalized to the maximum allowable weight for a contributing clock, are shown in Fig.~\ref{f.weights}. On MJD 56074, the weighting algorithm emphasized stable clocks, with a weight given by the inverse of the classical variance of the clock's frequency with respect to EAL over 30 days, $1/\sigma^2$, averaged over the previous 12 months. The maximum weight at that time was $2.5/N$, where $N$ is the number of clocks participating in that monthly determination.

The weighting algorithm was changed in 2014 to emphasize clocks that are predictable, assigning weight as the inverse of the variance of the measured frequency with respect to its predicted frequency using a model with an estimated drift. A typical model for expected clock reading $x$, at moment $t$ with respect to a superior reference is~\cite{10.1063/1.3681448} 
\begin{equation}
    x(t) = x_0+y_0t+y(t)t+\frac{1}{2}D(t)t^2+\sigma_x(t), \label{e.time_error}
\end{equation}
where $x_0$ is the initial time offset of the clock, $y_0$ is the initial frequency offset, $y(t)$ is the frequency and $D(t)$ is the rate of frequency drift at time $t$. The last term, $\sigma_x(t)$, characterizes stochastic fluctuations in the clock frequency/phase. In this scheme, weights are assigned in proportion to the square of the inverse clock deviation from the expected value above.

At the same time the weighting algorithm changed, the maximum allowable weight was increased to $4/N$~\cite{Panfilo_2014}. This resulted in a change in the relative emphasis of cesiums and masers in EAL. The maximum weight was again changed in 2022 to $6/N$. The change in weighting algorithm and in maximum weight can be seen at MJD 56684 and MJD 59849 in Fig.~\ref{f.weights} by the dashed vertical lines. 

Three of the four USNO rubidium fountains were assigned the maximum weight for each of the 144 months, while one was occasionally assigned a slightly lower weight. 
By MJD 60429, the four rubidium fountains were contributing more combined weight to UTC than the more than 200 high-performance cesium beams that report.

\section{Timing Stability, Timing Error and Holdover}

\subsection{TDEV}

\begin{figure}
\includegraphics[width=0.45\textwidth]{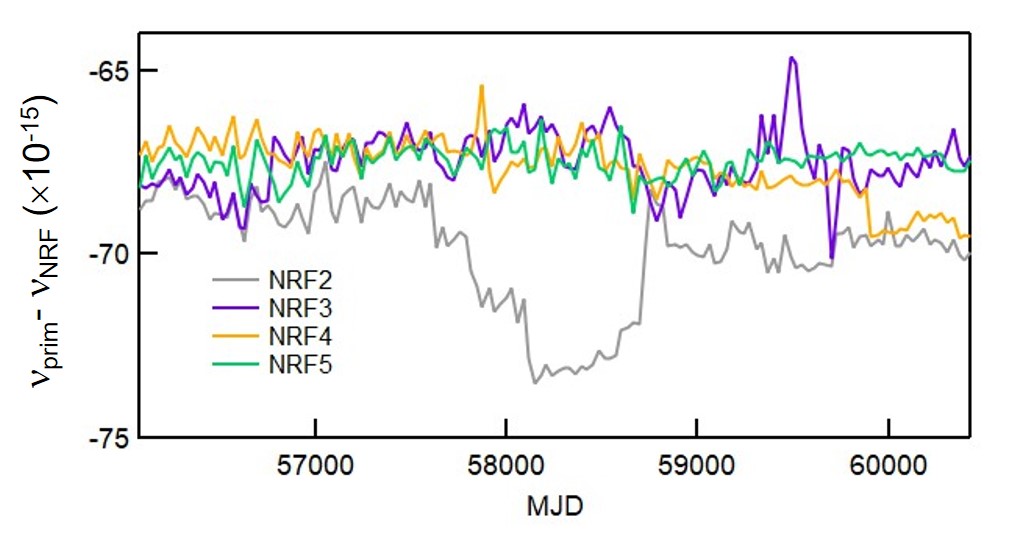}
\caption{Record of fountain frequencies with respect to a weighted average of the primary frequency standards over 12 years.} \label{f.fnt_rates}
\end{figure}

\begin{figure}
\includegraphics[width=0.45\textwidth]{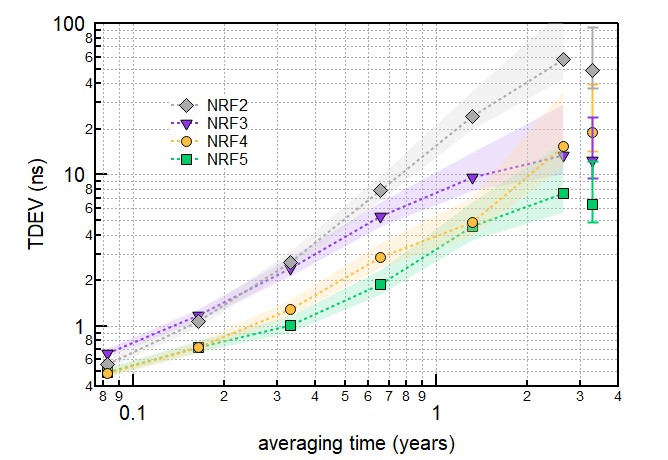}
\caption{TDEV for each of the four rubidium fountains using 12 years of BIPM data, $\nu_{\mathrm{prim}}-\nu_{\mathrm{NRF}}$.  Uncertainties are expressed by the shaded regions.  The last point shown with error bars for each clock is an evaluation of Total TDEV~\cite{717904}.} \label{f.tdev}
\end{figure}

Time deviation (TDEV), $\sigma_x(\tau)$, is the uncertainty associated with the measurement of a continuous time interval, $\tau$. Calculation of TDEV requires an uninterrupted frequency (or phase) record for a given clock. To get the best long term frequency reference to characterize the performance of the USNO rubidium fountains, we use their frequency records with respect to the primary standards that define the SI second. The frequencies published by the BIPM use TAI as the reference, i.e. $\nu_{\mathrm{TAI}}-\nu_{\mathrm{clock}}$. We adjust these using the published values of the frequency difference between TAI and the primary standards, $\nu_{\mathrm{prim}}-\nu_{\mathrm{TAI}}$, to get $\nu_{\mathrm{prim}}-\nu_{\mathrm{clock}}$.  The continuous frequency records for each USNO rubidium fountain measured against the primary standards over 12 years are shown in Fig.~\ref{f.fnt_rates}, and TDEVs for the USNO rubidium fountains
from the 12~year data record are plotted in Fig.~\ref{f.tdev}. The TDEV values remain below 100~ns at all averaging times out to 3 years, with the best performing fountain below 10~ns. 

\subsection{Timing Error}

The model used for weighting clocks, Eq.~(\ref{e.time_error}), predicts the time error accumulated over a particular interval. The offsets $x_0$ and $y_0$ can be reset or recalibrated during synchronization and syntonization events. Calibration of frequency and frequency drift can be used to project expected clock reading, so when in holdover (free running), the time of the superior reference can be predicted and systematic time error only accumulates due to imperfect calibration or changes in $y(t)$ and $D(t)$.  For the longest intervals, errors in calibration or changes in $D(t)$ have the most dramatic effect on time error.

\begin{figure}
\includegraphics[width=0.45\textwidth]{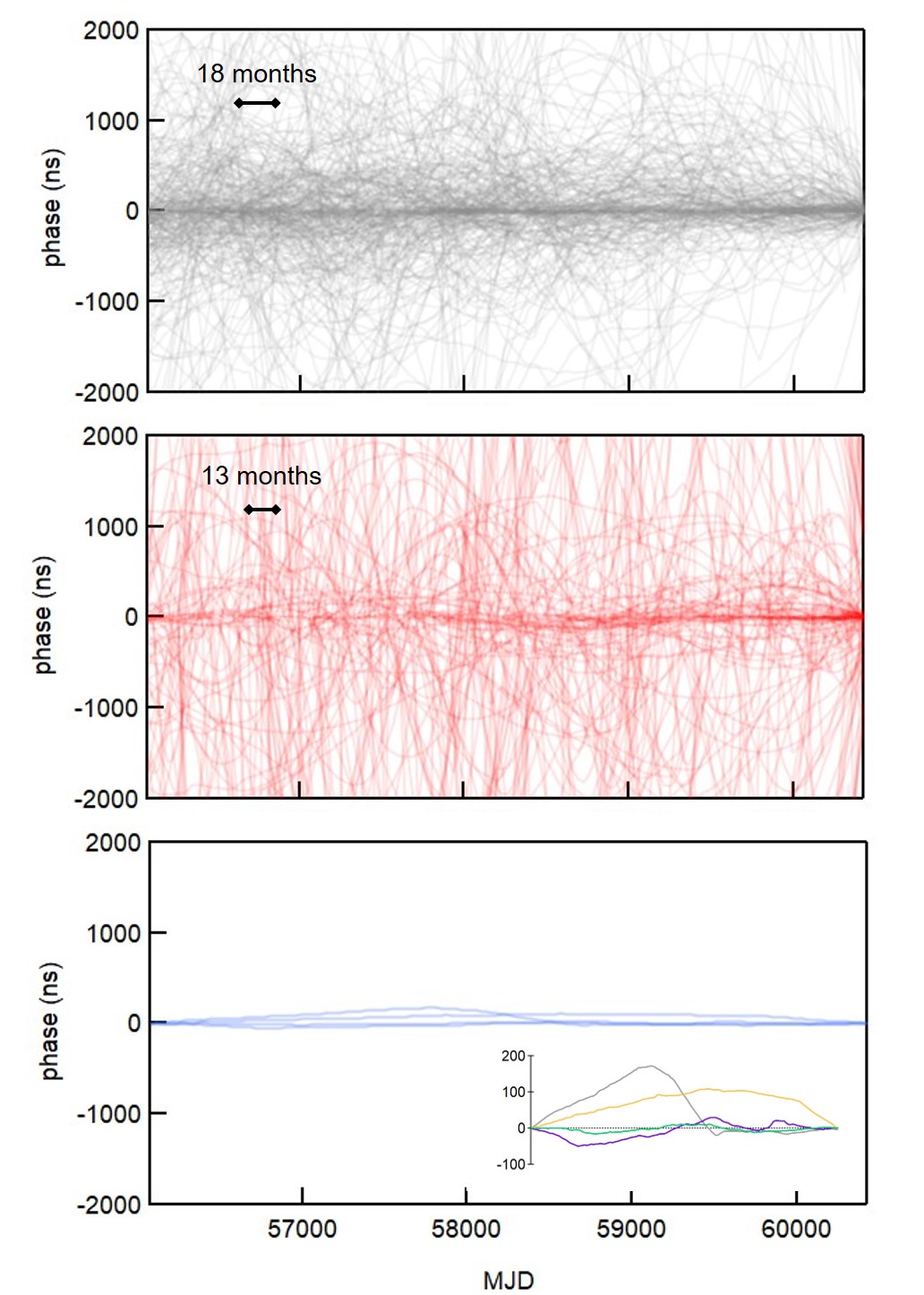}
\caption{Plots of residual phase difference with respect to TAI for each continuous reporting interval of cesium beams (top) and hydrogen masers (middle) reporting to the BIPM. 
The black horizontal line indicates the median length of continuous reporting periods for each type of clock. The bottom plot shows the residual phase for the USNO rubidium fountains with respect to the primary standards reporting to the BIPM; the continuous reporting period is the entire 144 months, so no indicating line is included. Inset: Rescaled plot of residual phase for the fountains, with individual fountains indicated with same color scheme as previous plots.} \label{phases_all}
\end{figure}

We can evaluate the maximum phase excursion of a clock with respect to the primary standards after removing a single frequency difference. For the USNO fountains, we integrate each frequency record to obtain the relative phase between the clock and the primary standards, and then remove a single frequency. The residual phase fluctuations exhibit peak-to-peak values over 12~years that range from 25~ns for NRF5 to 200~ns for NRF2.  Figure~\ref{phases_all} shows the residual phase difference after removing a single frequency offset for continuous reporting intervals for cesium beams, hydrogen masers, and the USNO rubidium fountains.  The same scaling is chosen for all three plots, even though many of the cesium beam and maser phase records extend beyond $\pm 2~\mu$s.

More relevant than removing a single frequency ex post facto, we consider an ``experiment'' where we assess how well fountains can holdover the phase of a timescale using the primary standards reporting to the BIPM as the stable frequency reference after a period of synchronization. Clocks reporting to the BIPM for contribution to UTC must wait through a 6 month probationary period before being included in the timescale. When the USNO rubidium fountains came online in March of 2011, MJD 55637, characterization was carried out locally for 8 months before the fountains were reported to the BIPM.  After the additional 6 month probationary period, the fountains were used in the generation of TAI starting on MJD 56074.

\begin{figure}
\includegraphics[width=0.45\textwidth]{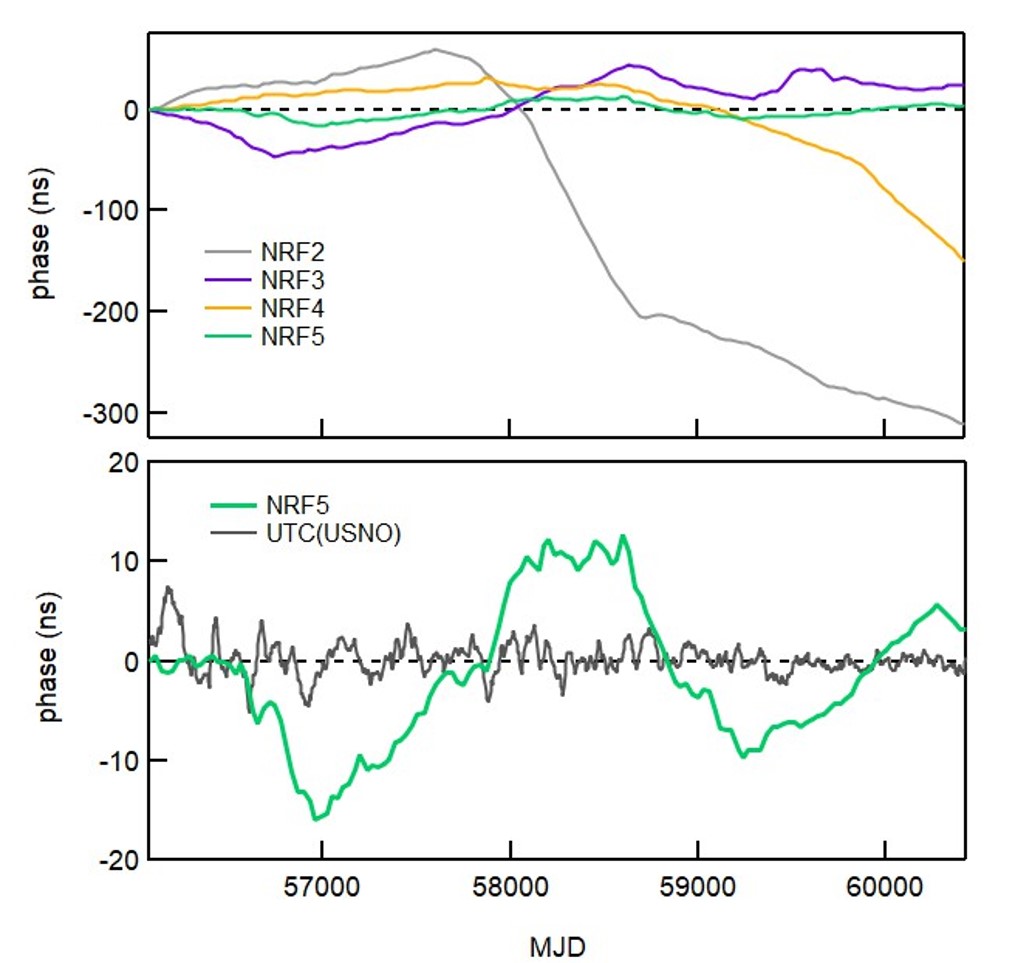}
\caption{Residual phase difference between each rubidium fountain and the primary standards after after ``synchronizing'' for 1~year (top). Emphasis of NRF5 from top plot, along with UTC(USNO) for comparison.} \label{f.phase}
\end{figure}

\begin{figure}
\includegraphics[width=0.45\textwidth]{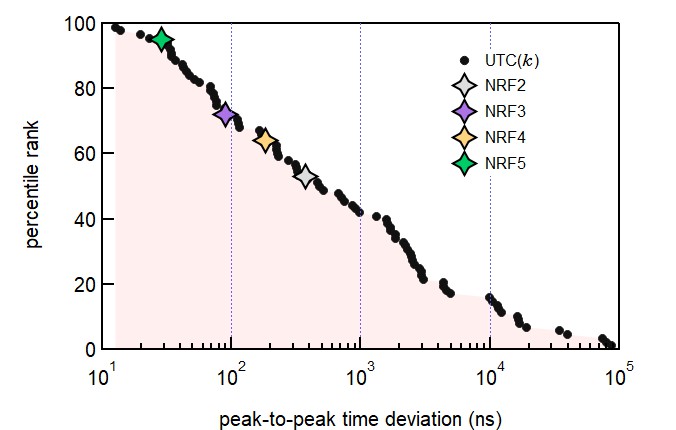}
\caption{Percentile rank of every UTC($k$) that reported between MJD 56074 and MJD 60429 in terms of peak-to-peak time deviation from UTC over that interval (black dots). As a comparison, each USNO fountain is shown where it would lie on this plot according to its peak-to-peak deviation from the primary standards over this same interval. Data for the peak-to-peak deviation of UTC($k$) with respect to UTC over this 12 year epoch is calculated from published reports of UTC($k$)-UTC in the Circular T~\cite{circulart}.} \label{f.utc_x}
\end{figure}

We use this $\sim 1$~year of fountain operation at USNO before the clocks were weighted in TAI as a synchronization period. Using the frequency of UTC(USNO) as a proxy for that of TAI,
we remove the frequency difference measured for 1~year from the 12-year record of $\nu_{\mathrm{TAI}}-\nu_{\mathrm{NRF}}$, add the difference $\nu_{\mathrm{prim}}-\nu_{\mathrm{TAI}}$, integrate the result and assess the phase residuals.  This produces the curves in Fig.~\ref{f.phase} (top plot). The fountain that has exhibited the most frequency instability over long times, NRF2, shows the worst holdover performance, with peak-to-peak deviation from the primary standards at 12~years of 350~ns. At the other extreme, NRF5 maintains the timescale within $\pm 14$~ns after 12~years (Fig.~\ref{f.phase} (bottom plot)).  As a comparison, of the more than 80 composite-clock, steered timescales UTC($k$), only 4 maintained a smaller maximum deviation from UTC over the same period than did NRF5, a single unsteered clock, with respect to the primary standards (Fig.~\ref{f.utc_x}).

\section{Conclusion}

Rubidium atomic fountains in use for 13 years operate as clocks as opposed to frequency standards and have contributed to international timescales for more than 12 years. A continuous phase record over this long epoch allows us to demonstrate timing capabilities over unprecedented intervals. Continuous operation over this period of time was facilitated by a well controlled environment as well as the availability of a maser for emergency holdover. With the upgrade to a telecom-based laser system, reliance on an emergency flywheel is no longer required. While the specific use of the fountains alleviates constraints on size, weight and power (SWaP) that are a concern for most commercial clocks, the performance demonstrated shows the long-term timing potential possible with cold-atom clocks with well-controlled systematic frequency shifts. 

Precise time, in particular UTC, is available globally via GNSS and to timing labs via the BIPM. Independence from these for long epochs requires reliable, long-term-stable and low (or constant) drift clocks such as the cold-atom clocks analyzed here. Fountains demonstrate the ability to holdover the best international timescales at the level of hundreds to tens of nanoseconds over more than a decade. Such independence is important for resiliency, and it may be be beneficial for a future lunar presence where time transfer with a superior standard is limited~\cite{Thompson_2023}. 

\section{Acknowledgements}

The fountains at USNO were designed and constructed over many years, with contributions from many people at USNO. This work is supported by Department of Navy Research, Development, Test and Evaluation funds provided for USNO’s Clock Development program.

\section{Appendix}

\subsection{Clock Types}

The BIPM categorizes clocks using a number of different codes, most of which are reserved for a variety of (commercial) cesium beam or hydrogen maser, accounting for almost all of the hundreds of clocks contributing each month.  On MJD 56074, only 6 clocks that reported were not categorized as a type of maser or (commercial) cesium, two laboratory cesium beams at the Physikalisch-Technische Bundesanstalt (PTB) and USNO's four rubidium fountains in Washington, DC.  These were designated type 9x for ``PRIMARY CLOCKS AND PROTOTYPES'' (92 for the cesium beams, 93 for the rubidium fountains). Prior to MJD 56074, only 4 other 9x type clocks had ever reported: a third cesium beam at PTB (type 92), two laboratory cesium beams at NRC, the National Research Council of Canada, (type 90), and a microwave mercury ion clock at AUS, the National Measurement Institute in Australia (type 99).

Over the 12 years of fountain reporting, the 9x clock codes evolved to two specific codes, 92, for ``GROUND-STATE HYPERFINE TRANSITION OF 133 Cs'', and 93, for ``GROUND-STATE HYPERFINE TRANSITION OF 87 Rb''. Many more codes were added during that time to account for future optical clocks.  Yet, on MJD 60429, there were still only 8 clocks not categorized as (commercial) cesiums or masers, the two PTB cesium beams, the four USNO rubidium fountains, 1 rubidium fountain at the National Time Service Center (NTSC) in China~\cite{9933887}, and a cold rubidium clock at IT (INRIM) (type 42). %Clock type 70 reported 2 times, never weighted. 
Of the 1,229 clock reports since inception of TAI through MJD 56074, the only 93 clocks have been USNO fountains (4 fountains, 144 months each) and the NSTC fountain (1 fountain, 12 months).  The rubidium fountain results and analysis in this work address only the four fountains at USNO in Washington, DC. Similarly, the only commercial cesium clocks included in our analysis are type 35, MICROCHIP 5071A/5071B HIGH PERFORMANCE TUBE, and the only masers included are type 40, UNSPECIFIED HYDROGEN MASER.  

\section*{References}

%\bibliography{holdover}


\end{document}